\def\draftversion{false}
\RequirePackage{ifthen}
\ifthenelse{\equal{\draftversion}{true}}{
  \documentclass[floatfix,galley,prb]{revtex4}
}{
  \documentclass[floatfix, twocolumn, prb]{revtex4}
}

\usepackage{color}
\usepackage{multirow}
\usepackage{epsfig}
\usepackage{graphicx}
\usepackage{amsmath}
\usepackage{amssymb}
\usepackage{bm}
\usepackage{dcolumn}
\usepackage{braket}
\usepackage{bbold}

\begin{document}

\title{Quantum anomalous Hall phase in (001) double-perovskite monolayers \\
via intersite spin-orbit coupling}

\author{Hongbin~Zhang$^1$}
\email[corresp.\ author: ]{hzhang@physics.rutgers.edu}
\author{Huaqing~Huang$^{1,2}$}
\author{Kristjan~Haule$^1$}
\author{David~Vanderbilt$^1$}
\affiliation{$^1$Department of Physics and Astronomy, Rutgers
University, Piscataway, USA \\
$^2$Department of Physics, Tsinghua University, Beijing 100084, China}

\date{Jun. 10, 2014}

\begin{abstract}
Using tight-binding models and first-principles calculations,
we demonstrate the possibility to achieve a quantum anomalous Hall
(QAH) phase on a two-dimensional square lattice, which can be realized
in monolayers of double perovskites.  We show that effective
intersite spin-orbit coupling between $e_g$ orbitals
can be induced perturbatively, giving rise to a QAH state.  
Moreover, the effective spin-orbit coupling can be enhanced by
octahedral rotations. Based on first-principles calculations,
we propose that this type of QAH state could be realized in La$_2$MnIrO$_6$
monolayers, with the size of the gap as large as $26$\,meV in
the ideal case.  We observe that the electronic structure is
sensitive to structural distortions, and that an enhanced Hubbard
$U$ tends to stabilize the nontrivial gap.
\end{abstract}

\maketitle

%-------- MARGIN COMMENTS --------------
\def\scr{\scriptsize}
\ifthenelse{\equal{\draftversion}{true}}{
  \marginparwidth 2.7in
  \marginparsep 0.5in
  \newcounter{comm} % counter for commentaries
  % increase counter
  \def\commnext{\stepcounter{comm}}
  % commentary in text
  \def\commtext{{\bf\color{blue}[\arabic{comm}]}}
  % commentary in margin
  \def\commmar{{\bf\color{blue}[\arabic{comm}]}}
  % comment commands for all authors
  \def\dvm#1{\commnext\marginpar{\small DV\commmar: #1}\commtext}
  \def\hzm#1{\commnext\marginpar{\small HZ\commmar: #1}\commtext}
  \def\khm#1{\commnext\marginpar{\small KH\commmar: #1}\commtext}
}{
  \def\dvm#1{}
  \def\hzm#1{}
  \def\khm#1{}
}
%----------------------------------

%-------- COLORS --------------
\def\Red#1{\textcolor{red}{#1}}
\def\Blue#1{\textcolor{blue}{[#1]}}
\def\Magenta#1{\textcolor{magenta}{#1}}
%----------------------------------

%%%%%%%%%%%%%%%%%%%%%%%%%%%%%%%%%%%%%%%%%%%%%%%%%%
\section{Introduction}
%%%%%%%%%%%%%%%%%%%%%%%%%%%%%%%%%%%%%%%%%%%%%%%%%%

The quantum anomalous Hall (QAH) effect has drawn intensive attention
recently, in part due to the dissipationless transport that can take
place in the spin-polarized edge states, which are
topologically protected against perturbative disorder.
A generic model to achieve the QAH phase was first proposed by Haldane
on the honeycomb
lattice,~\cite{Haldane:1988} where complex
hoppings between next-nearest neighbors (NNNs) play a crucial role.
Several systems have been proposed to host such nontrivial
topological phases, such as magnetically doped topological
insulators~\cite{Liu:2008, Yu:2010} and honeycomb lattices formed by
transition-metal or heavy-metal ions.~\cite{Qiao:2010, Zhang:2012, Hu:2012,
Garrity:2013, Wang:2013}
For most of these systems, the occurrence of the QAH phase relies on the
honeycomb lattice, and the topological properties are usually carried by
the $sp$ bands.
Meanwhile, spontaneous time-reversal symmetry breaking is usually
induced by doping with magnetic ions or via a magnetic proximity effect.
These two limitations greatly reduce the range of available candidate systems
to search for the occurrence of a QAH state.
In Cr-doped (Bi$_{1-x}$Sb$_x$)$_2$Te$_3$, for example, where the QAH phase has first
been observed experimentally,~\cite{Chang:2013} the QAH effect is only
observable below about $30$\,mK, due to the small exchange splittings induced
by Cr doping.

In their seminal work, Xiao~{\it et al.}~proposed that in (111) superlattices
of perovskite transition-metal oxides (TMOs), various topological phases
can be obtained.~\cite{Xiao:2011}
For TMOs with partially occupied $d$ shells, magnetism is relatively
easy to obtain because the $d$ electrons are more localized than the
$sp$ electrons.
Furthermore, electronic correlations are usually significant in
TMOs with localized $d$ electrons, and there is the possibility that
nontrivial topological phases can
develop by spontaneous symmetry breaking~\cite{Raghu:2008,
Sun:2009, Maciejko:2014} with a dynamically generated spin-orbit
coupling (SOC).~\cite{Wu:2004}  It has even been theoretically argued
that nontrivial topological
phases can be realized in (111) TMO heterostructures without
considering atomic SOC.~\cite{Yang:2011,Rueegg:2011}
In all these proposals, the underlying honeycomb lattice facilitates the
appearance of a topological phase.
Unfortunately, it is difficult to synthesize (111) TMO superlattices
experimentally with good atomic precision, although there has been some
recent experimental progress in this direction.~\cite{Gibert:2012}

The presence of a honeycomb lattice is not, however, a necessary condition for
the occurrence of the QAH effect.
For instance, topologically nontrivial phases can be obtained on
square lattices with well designed nearest-neighbor (NN) and NNN
hoppings.~\cite{Wang:2012, Yang:2012, Guo:2013}
Recently, three
proposals have been put forward to achieve the QAH effect in more
realistic systems based on square-lattice symmetry,~{\it i.e.},~superlattices of
CdO/EuO~\cite{Zhang:2013} and GdN/EuO~\cite{Garrity:2014} with the rocksalt 
structure and CrO$_2$/TiO$_2$ with the rutile structure.~\cite{Cai:2013}
For the latter case, the relevant bands are the $t_{2g}$ states of Cr;
while these states show large exchange splittings, the topological gap is only 
about $4$\,meV due to the small strength of the on-site atomic SOC of Cr atoms.

In this work, we demonstrate the possibility of achieving a nontrivial QAH
phase in (001)-oriented double-perovskite monolayers.
Using a two-band model for $e_g$ orbitals on a square lattice, we
show that complex effective intersite hoppings between two $e_g$ orbitals
can be induced perturbatively by the atomic SOC,
giving rise to a QAH state.
Based on first-principles calculations, we further show that
such a model can be realized in checkerboard La$_2$MnIrO$_6$ (LMIO)
monolayers (MLs) embedded in a non-magnetic insulating host such as
LaAlO$_3$ (LAO).
The magnitude of the topological gap in the ideal case can be
as large as $26$\,meV.
The advantage of such a system is that (001) superlattices of
perovskite compounds are well studied and can be synthesized
with good atomic precision, resulting in controlled  structural properties.
Moreover, given the abundance of physical properties in perovskite
TMO superlattices, including high-T$_c$ superconductivity,~\cite{Logvenov:2009} 
the QAH phase realized in (001) perovskite
superlattices can also be integrated more easily with other functional
oxides to achieve new physical properties.

The manuscript is organized as follows.  In Sec.~II we present the tight-binding
model for half-filled $e_g$ states on a square lattice. We demonstrate how
the effective SOC can be induced following standard perturbation
theory, focusing on the role of octahedral rotations. Detailed symmetry analysis
is given to understand how the nontrivial topological phase develops.
Our first-principles results are shown in Sec.~III, where the model arguments 
in Sec.~II are verified by considering a hypothetical structure. 
The effects of structural relaxations are then studied in detail and it is shown that 
epitaxial strain can be used to tune the LMIO monolayers close to the critical
region where a nontrivial QAH state exists. 
%%%%%%%%%%%%%%%%%%%%%%%%%%%%%%%%%%%%%%%%%%%%%%%%%%
\section{two-band model}
%%%%%%%%%%%%%%%%%%%%%%%%%%%%%%%%%%%%%%%%%%%%%%%%%%

%------------------------------
%Fig1: sketch of the model
%------------------------------
\begin{figure}[t]
\includegraphics[width=9.0cm]{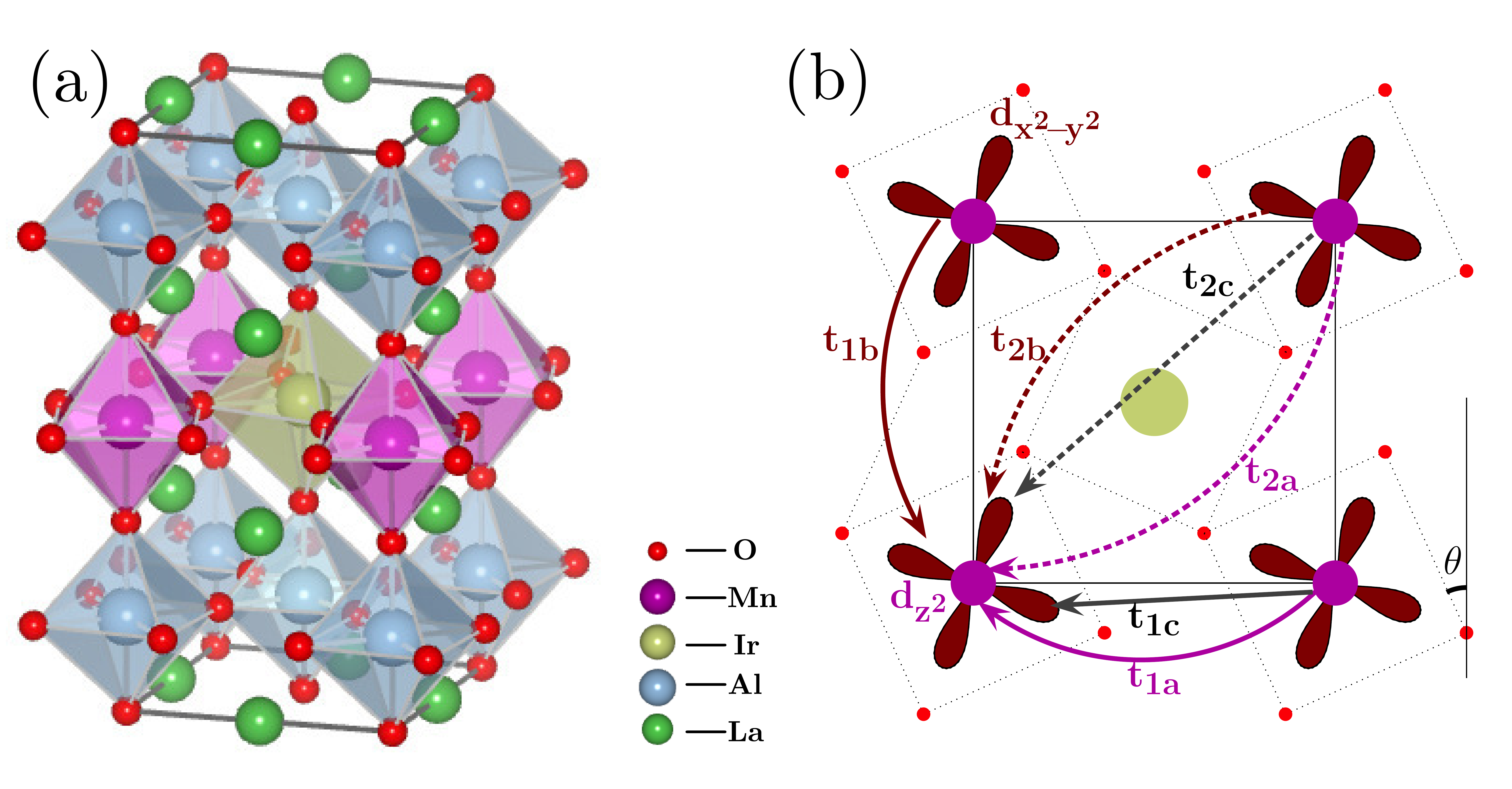}
\caption{(Color online)
(a) Oblique view of the crystal structure of an La$_2$MnIrO$_6$
(LMIO) monolayer (ML) sandwiched between LaAlO$_3$ layers.
(b) Sketch defining the parameters of the tight-binding model of
Eq.~(\ref{eq:model}) that describes the LMIO ML.
Only the local $e_g$ orbitals on Mn atoms (purple, at origin) are shown;
orbitals on Ir (brown, at the center) are suppressed.
The intersite hoppings between Mn $e_g$ orbitals on nearest neighbors
($t_{1i}$) and next-nearest neighbors ($t_{2i}$) are shown by
arrows (see main text for details).
The octahedral rotations, denoted by angle $\theta$ in (b), are
exaggerated for clarity of illustration.
}\label{fig:model}
\end{figure}

Our tight-binding model simulates a double-perovskite ML with
checkerboard ordering of the two sublattices, either isolated in vacuum or
embedded in an inert (wide-gap non-magnetic) perovskite host.
To be specific, we consider a case in which one sublattice is
populated with ions having large exchange splittings, typically
high-spin $3d$ transition-metal ions, while ions with filled
({\it e.g.}, $t_{2g}$) subshells, preferably with large on-site
SOC, are located on the other sublattice.
In this work we consider the combination of Mn$^{3+}$ and Ir$^{3+}$
ions as an example, as it will turn out to be a promising candidate
based on our first-principles calculations in Sec.~III.

The crystal structure of such an LMIO monolayer sandwiched between
LAO layers is shown in Fig.~\ref{fig:model}(a).
The corresponding tight-binding model for the two $e_g$ orbitals on each
Mn site can be expressed in the local  $(d_{z^2}, d_{x^2\!-\!y^2})$ basis as
\def\kk{\mathbf{k}}
\begin{widetext}
\begin{equation}
H\!=\!
\begin{pmatrix}
  t_{1a}\,f_1(\kk)\!+\!t_{2a}\,f_2(\kk) &
  (t_{1c}\!-\!i\lambda^{(1)})\,g_1(\kk)\!-\!(t_{2c}\!-\!i\lambda^{(2)})\,g_2(\kk) \\
  \textrm{c.c.} &
  \Delta\!+\!t_{1b}\,f_1(\kk)\!+\!t_{2b}\,f_2(\kk)
\end{pmatrix}
\label{eq:model}
\end{equation}
\end{widetext}
where $f_1(\kk)\!=\!\cos k_x\!+\!\cos k_y$, $g_1(\kk)\!=\!\cos
k_x\!-\!\cos k_y$, $f_2(\kk)\!=\!2\cos k_x\cos k_y$, and
$g_2(\kk)\!=\!2\sin k_x\sin k_y$.  The model is parametrized by
the difference $\Delta$ between the on-site energies of
$d_{x^2-y^2}$ and $d_{z^2}$ orbitals,
the NN hoppings $t_{1i}$, the NNN hoppings $t_{2i}$,
and the effective SOC parameters $\lambda^{(1)}$
and $\lambda^{(2)}$, which respectively denote the NN and NNN
couplings between $d_{z^2}$ and $d_{x^2\!-\!y^2}$ orbitals induced
perturbatively as explained below.
For the hoppings $t_{1i}$ and $t_{2i}$, $i\!=\!a$ or $b$ refers
to the like-orbital hopping between $d_{z^2}$ or $d_{x^2\!-\!y^2}$
orbitals
respectively, while $i\!=\!c$ denotes the unlike-orbital
hopping between $d_{z^2}$ and $d_{x^2\!-\!y^2}$ orbitals.
Note that these are all ``effective hoppings'' in the sense that
the oxygen and iridium orbitals are regarded as having been
integrated out.

%------------------------------
% lambda^1 between NN
%------------------------------
The complex hopping terms $i\lambda^{(1)}$ and $i\lambda^{(2)}$ in 
Eq.~(\ref{eq:model})
between $d_{z^2}$ and $d_{x^2\!-\!y^2}$ orbitals can be induced
by considering perturbative processes involving SOC.
When there is no rotation of the transition-metal-oxygen octahedra, 
the $i\lambda^{(1)}$ term arises following
\begin{equation}
i\lambda^{(1)}=\frac{\braket{d^\text{Mn}_{z^2}\!\mid \hat{H}
  \!\mid\!d^\text{Mn}_{xy}}\braket{d^\text{Mn}_{xy}
  \!\mid\!\xi^\text{Mn}\hat{\mathbf{L}}\cdot\hat{\mathbf{S}}\!\mid\!
  d^\text{Mn}_{x^2\!-\!y^2}}}{E_{e_g}^\text{Mn}-E_{t_{2g}}^\text{Mn}}
\end{equation}
where $\hat{\mathbf{L}}$ ($\hat{\mathbf{S}}$) is the orbital
(spin) angular-momentum operator,
$\xi^\text{Mn}$ is the strength of the atomic SOC on Mn, and
$\hat{H}$ denotes direct hybridization between $d_{xy}$ and
$d_{z^2}$ orbitals located on NN Mn sites.
$E_{t_{2g}}^\text{Mn}$ and $E_{e_g}^\text{Mn}$ denote the 
on-site energies of the $t_{2g}$ and $e_g$ subshells
on the Mn ions in the cubic crystal field.
In a general case when the in-plane rotation angle $\theta$ 
(cf.~Fig.~\ref{fig:model}(b)) is nonzero, 
it can be shown that $i\lambda^{(1)}\!\propto\!
i\xi^\text{Mn}\sin(2\theta^\text{Mn})\!=\!i\xi^\text{Mn}\cos(2\theta)$, where
$\theta^\text{Mn}$ denotes the rotation angle of the MnO$_6$ octahedra
and $\theta^\text{Mn}\!=\!\theta\!+\!45^\circ$.  
That is, $i\lambda^{(1)}$ is an even function of the rotation angle $\theta$.
We note that second-order
processes involving Ir $t_{2g}$ orbitals can also lead to an effective
SOC between the $e_g$ orbitals located on NN Mn sites, 
but the two most obvious contributions, corresponding to hopping 
via the two Ir atoms adjacent to a given Mn-Mn bond, tend to cancel 
one another.

%------------------------------
% lambda^2 between NNN
%------------------------------
A finite rotation angle $\theta$ leads to a nonzero $i\lambda^{(2)}$ term 
between $e_g$ orbitals located on NNN Mn sites.
It arises following
\begin{widetext}
\begin{equation}
i\lambda^{(2)}=\frac{\braket{d^\text{Mn}_{z^2}\!\mid \hat{H}^\prime\!\mid\!d^\text{Ir}_{xy}}
   \braket{d^\text{Ir}_{xy}\!\mid\!\xi^\text{Ir}\hat{\mathbf{L}}\cdot\hat{\mathbf{S}}\!\mid\!d^\text{Ir}_{x^2\!-\!y^2}}
   \braket{d^\text{Ir}_{x^2\!-\!y^2}\!\mid\! \hat{H}^\prime\!\mid\!d^\text{Mn}_{x^2\!-\!y^2}}}
   {(E^\text{Ir}_{e_g}\!-\!E^\text{Ir}_{t_{2g}})(E^\text{Mn}_{e_g}-E^\text{Ir}_{e_g})}
\end{equation}
\end{widetext}
where $\xi^\text{Ir}$ denotes the strength of the atomic SOC on Ir sites,
$E^\text{Ir}_{t_{2g}}$ and $E^\text{Ir}_{e_g}$ are the on-site 
energies of the Ir $t_{2g}$ and $e_g$ orbitals, and 
$\hat{H}^\prime$ denotes the direct hybridization between orbitals on 
Mn and Ir atoms. 
Similar virtual transitions involving coupling of the $d^\text{Mn}_{x^2-y^2}$ 
orbital of Mn to the $d^\text{Ir}_{xy}$ orbital of Ir also lead to 
nonzero contributions.
The resulting total $i\lambda^{(2)}\!\propto\!i\xi^\text{Ir}\sin(2\theta)$, with $\theta$ 
the octahedral rotation angle. That is, $i\lambda^{(2)}$ is an odd function of
$\theta$.
Furthermore, the magnitude of $\lambda^{(2)}$ is determined by the strength
of the atomic SOC of the Ir atoms.
We observe that for the LMIO monolayers considered
in this work, the magnitude of $\lambda^{(2)}$ is about one order of 
magnitude larger than that of 
$\lambda^{(1)}$, due to the much stronger atomic SOC of Ir 
($\sim\!0.5$\,eV) compared to that of Mn ($\sim\!0.05$\,eV).

%------------------------------
%decomposition  of the 2X2 Hamiltonian
%------------------------------
The two-band model of Eq.~(\ref{eq:model}) can be solved analytically by
decomposing the Hamiltonian as $H\!=\!\sigma_0 h_0\!+\!\sum_{i\!=\!1}^3 h_i\sigma_i$ where
$\sigma_0$ is the unit matrix and $\sigma_i$ are the Pauli matrices.
The Berry curvature for the lower-lying band can be obtained explicitly as
\begin{equation}
\Omega=-\frac{2}{h}\frac{\epsilon_{ijk}\,h_ih_{jx}h_{ky}}
   {(E_+\!-\!E_-)^2}
\label{eq:anasolv}
\end{equation}
where $\epsilon_{ijk}$ is the Levi-Civita symbol,
$h\!=\!\sqrt{h_1^2\!+\!h_2^2\!+\!h_3^2}$, $h_{i\alpha}\!=\!\partial h_i/\partial
k_\alpha$ ($\alpha\!=\!x,y$), and $E_\pm\!=\!h_0\pm h$ are the energy eigenvalues for
the higher/lower bands.  In our case,
\begin{equation}
\begin{split}
h_0&=\left[\Delta\!+\!(t_{1a}+t_{1b})f_1(\kk)\!+\!(t_{2a}+t_{2b})f_2(\kk)\right]/2\,,\\
h_1&=t_{1c}g_1(\kk)-t_{2c}g_2(\kk)\,,\\
h_2&=\lambda^{(1)} g_1(\kk)-\lambda^{(2)} g_2(\kk)\,,\\
h_3&=\left[-\Delta\!+\!(t_{1a}-t_{1b})f_1(\kk)\!+\!(t_{2a}-t_{2b})f_2(\kk)\right]/2\,. \\
\end{split}
\label{eq:decompose}
\end{equation}
\normalsize

\begin{figure}[t]
\includegraphics[width=8.0cm]{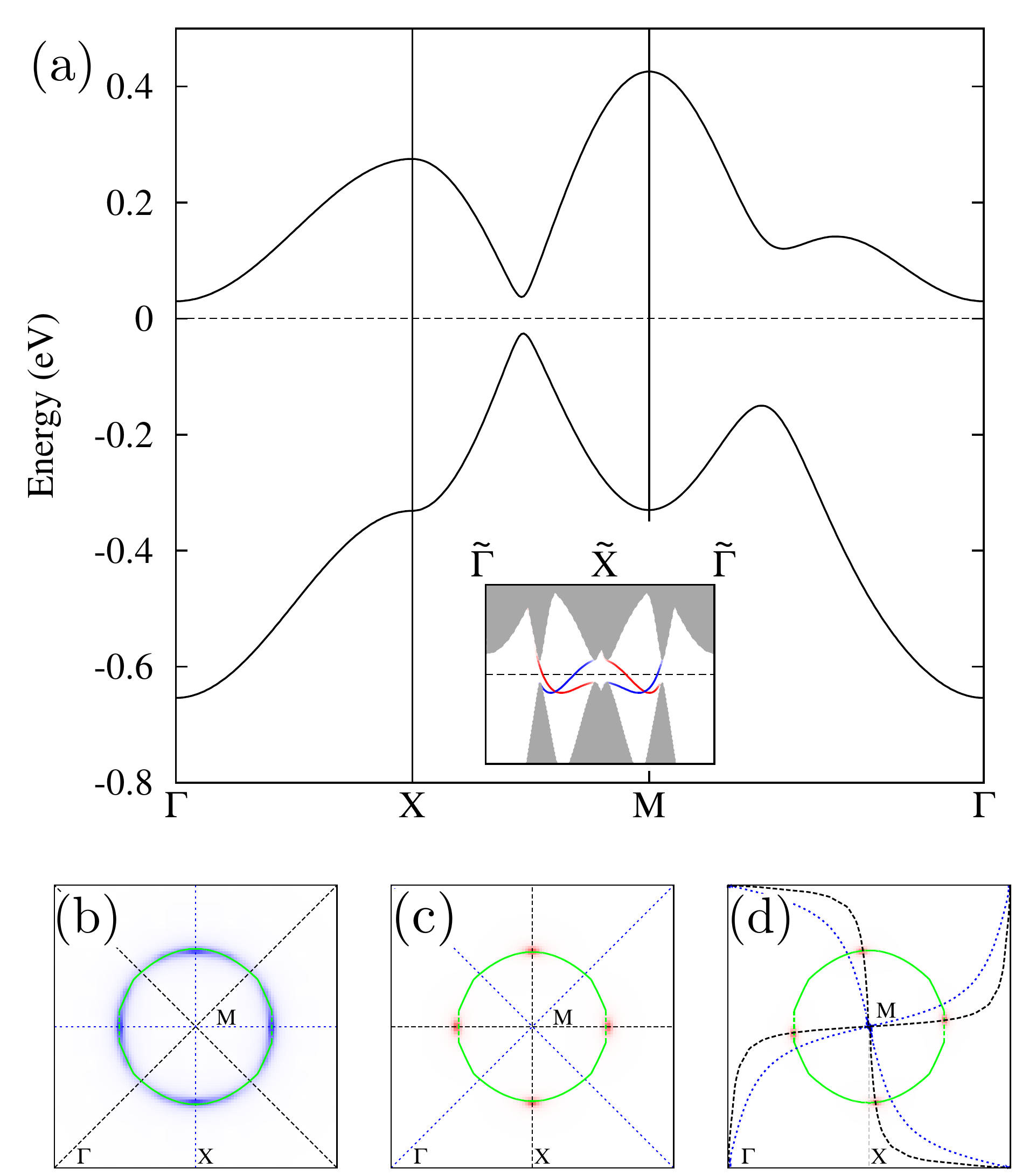}
\caption{(Color online)
Electronic structure of the tight-binding model of Eq.~(\ref{eq:model})
with parameters as given in the text.
(a) Band structure along the high-symmetry $\Gamma$-X-M-$\Gamma$ $k$-path.
Inset shows the projected bulk band structure (shaded region) and
edge states for an 80-unit-cell-wide ribbon in the reduced 1D BZ 
along $\tilde{\Gamma}$-$\tilde{\text{X}}$-$\tilde{\Gamma}$,
color-coded to distinguish the contributions from the two edges;
dashed line denotes $E_F$.
(b-d) Color map showing the distribution of negative (blue) and positive 
(red) Berry curvature in the BZ for three cases:
(b) without octahedral rotations ($t_{2c}$ and $\lambda^{(1)}$ nonzero);
(c) with octahedral rotations ($t_{1c}$ and $\lambda^{(2)}$ nonzero) but
$t_{2c}$ and $\lambda^{(1)}$ artificially set to zero; and
(d) with octahedral rotations, including all four terms.
For guidance, nodes of $h_1$, $h_2$, and $h_0\!+\!h_3$ in Eq.~(\ref{eq:decompose})
are shown as dotted blue, dashed black, and solid green curves, respectively.
}\label{fig:modelb}
\end{figure}

%------------------------------
% model parameters
%------------------------------
The band structure obtained from a model of this form is presented
in Fig.~\ref{fig:modelb}(a), and the regions of strong Berry
curvature, corresponding to small direct gaps, are shown as
the (blue/red) shaded regions in Fig.~\ref{fig:modelb}(b-d).
The diagonal hopping parameters for the plots were obtained by fitting to
the first-principles band structure of Fig.~\ref{fig:bands}(b),
yielding $t_{1a}\!=\!-0.27$\,eV, $t_{1b}\!=\!0.09$\,eV, $t_{2a}\!=\!0.05$\,eV,
$t_{2b}\!=\!-0.105$\,eV, and $\Delta\!=\!0.28$\,eV.
The off-diagonal terms were set to $t_{1c}\!=\!0.02$\,eV,
$t_{2c}\!=\!0.02$\,eV, $\lambda^{(1)}\!=\!0.02$\,eV, and $\lambda^{(2)}\!=\!0.08$\,eV.  
The inset of Fig.~\ref{fig:modelb}(a) shows the computed edge states
for an $80$-unit-cell-wide ribbon cut from this model, providing
the first evidence that the model exhibits a non-trivial topology.

%------------------------------
%Chern #
%------------------------------

To understand how these features of the band structure come about,
it is useful to return to Eqs.~(\ref{eq:anasolv}-\ref{eq:decompose}).
Note that $h_1$ and $h_2$ both have to be present in order to obtain a
nonzero Berry curvature.
Actually we find that nonzero $t_{2c}$ and $\lambda^{(1)}$ 
or nonzero $t_{1c}$ and $\lambda^{(2)}$ can both lead to
nontrivial topological phases, corresponding to the case without
rotations and the case with only terms induced by rotations,
respectively. Interestingly, the Chern numbers are of opposite
sign for the two cases. Due to the much larger magnitude of 
$\lambda^{(2)}$, the Chern number of the system is determined
in practice by the combination of $t_{1c}$ and $\lambda^{(2)}$.

%---------------------
% no rotation
%---------------------
Consider first
the case that the octahedral rotation angle $\theta$ vanishes.
Then $t_{1c}\!=\!0$ because symmetry prevents any
direct hybridization of $d_{z^2}$ and $d_{x^2-y^2}$ orbitals
on NN Mn sites.
In this case, $\lambda^{(2)}\!=\!0$ as well because the hybridization of
$d_{z^2}$ and $d_{x^2-y^2}$ orbitals on NNN Mn sites via Ir
atoms is forbidden.
In fact, we observed that both $t_{1c}$ and $\lambda^{(2)}$ is
proportional to $\sin{2\theta}$, as explained above for $\lambda^{(2)}$.
That is, without octahedral rotations only $t_{2c}$ and $\lambda^{(1)}$
in the off-diagonal terms of Eq.~(\ref{eq:model}) are nonzero.
Examining Eq.~(\ref{eq:model}) reveals
that without the off-diagonal terms proportional to $\lambda^{(1)}$
and $t_{2c}$, the eigenvalues of the Hamiltonian are degenerate
wherever $h_3\!=\!0$, which turns out to be
a loop centered at the M point in the Brillouin zone (BZ)
as shown by the green lines in Fig.~\ref{fig:modelb}(b).
This reflects the fact that neither $f_1(\mathbf{k})$ nor
$f_2(\mathbf{k})$ vanishes in the vicinity of M.
By contrast, $g_1(\mathbf{k})$ vanishes along the $\Gamma$-M
lines, while $g_2(\mathbf{k})$ vanishes along the X-M lines,
as indicated by the (black) dashed and (blue) dotted lines in
Fig.~\ref{fig:modelb}(b) respectively.
Thus, the loop of degeneracy is reduced to four points (Dirac
nodes) located on the X-M directions if $\lambda^{(1)}$ is turned
on, or on the $\Gamma$-M lines if $t_{2c}$ is turned on.

When both $\lambda^{(1)}$ and $t_{2c}$
are nonzero, the energy spectrum of Eq.~(\ref{eq:model}) is
fully gapped, leaving concentrations of Berry curvature in the
regions of the BZ where the gap is small, as shown by the
(blue) shading in Fig.~\ref{fig:modelb}(b).
Since the magnitude of $\lambda^{(1)}$ is comparable to that
of $t_{2c}$, the distribution of the Berry curvature is quite smeared.
The resulting total Chern number is $C\!=\!-2$
after integrating the Berry curvature over the whole BZ,
indicating that a QAH state has been formed.  This is
confirmed by our numerical calculation of the anomalous Hall
conductivity (AHC), which we find to be equal to $-2$\,$e^2/h$.

%------------------------------
%Fig2: band structures of the model
%------------------------------

%--------------------
% with only rotation
%--------------------
Switching on the octahedral rotations
modifies the hybridization between $d$ orbitals of Mn and Ir atoms,
renormalizing all the hopping parameters in Eq.~(\ref{eq:model}).
However, we find that the most important changes arise from the
fact that $t_{1c}$ and $\lambda^{(2)}$ adopt nonzero values when $\theta\!\ne\!0$.
To characterize the influence of these two terms, we first consider
an artificial situation in which $t_{1c}$ and $\lambda^{(2)}$ are turned
on while $t_{2c}$ and $\lambda^{(1)}$ are turned off, with all other
parameters kept fixed at their previous values.
As shown in Fig.~\ref{fig:modelb}(c), the sharpest concentrations of 
Berry curvature are found in monopole-like peaks; these lie on the X-M lines
because $|t_{1c}| <|\lambda^{(2)}|$. 
The Chern number that results by integrating the Berry curvatures over the
BZ is now $C\!=\!2$, opposite to the case when $t_{2c}$ and $\lambda^{(1)}$
are nonzero. This sign reversal results from the fact that the
$\lambda^{(1)}$ and $\lambda^{(2)}$ terms are of opposite sign (cf.~Eq.~\ref{eq:model}).

%----------------
% nonrot + rot = total
%----------------

Fig.~\ref{fig:modelb}(d) shows the distribution of the Berry curvature and the
corresponding nodes of the $h_1$, $h_2$, and $h_0+h_3$ defined in Eq.~(5)
when the full Hamiltonian of Eq.~(\ref{eq:model}) is considered, including 
both the preexisting interactions and those induced by rotations.
Since $\lambda^{(1)}$ and $\lambda^{(2)}$ have different $k$-dependences as
given in Eq.~(\ref{eq:model}), the nodes of $h_2$ shift slightly in the
counterclockwise direction relative to the case of Fig.~\ref{fig:modelb}(c).
There is also a shift (larger and in the opposite direction) in the nodes
of $h_1$ arising from the competition between $t_{1c}$ and $t_{2c}$.
However, because of the large magnitude of $\lambda^{(2)}$, the avoided
crossings follow the nodes of $h_2$. The result is that the
four avoided-crossing points of the full Hamiltonian are rotated slightly
in the counterclockwise direction around M compared to Fig.~\ref{fig:modelb}(c). 
Moreover, the topological properties of our model are dominated by the
combination of $\lambda^{(2)}$ and $t_{1c}$, so that the Chern number
is $C\!=\!2$. That is, the octahedral rotations induce a topological phase
transition where the Chern number changes from $-2$ to $2$.
The topological non-triviality is also confirmed by an explicit calculation of the 
edge states of a one-dimensional ribbon as shown in the inset of
Fig.~\ref{fig:modelb}(a). It is evident that two edge states with the same
group velocity are located on one edge, while another two edge states
with the opposite group velocity are located on the other edge.

%---------------
% Chern = 2
%----------------
The fact that the absolute value of the Chern number is two, and not one, 
can be understood in several ways.
For example, we note that in the limit of a very small coefficient
of $g_2$, so that the four Dirac points are just barely opened,
each should carry a Berry phase of $\pm\pi$ (as is standard
for a simple avoided Dirac crossing); four of them add to
$\pm4\pi$, suggesting $C\!=\!\pm2$.
Another approach is to consider the limit that the four degeneracy
points shrink to the M point and merge.
Specifically, we find that by tuning $\Delta$, the difference between
on-site energies of valence $d_{z^2}$ and conduction $d_{x^2-y^2}$
orbitals in the tight-binding model, the
four avoided-crossings shrink to a singular point with \textit{quadratic}
dispersion at M when $\Delta_{c}\!=\!1.034$\,eV, and the gap reopens
in a normal $C\!=\!0$ phase for $\Delta\!>\!\Delta_{c}$.
As pointed out in Ref.~\onlinecite{Bellissard:1995}, the Chern
number transfer should always be $\Delta C\!=\!2$ in the case of
a critical quadratic band touching.  This can also be understood
based on the symmetry of the orbitals.  At the M point, $d_{z^2}$
and $d_{x^2-y^2}$ states are both eigenstates of the $C_4$
symmetry operator, but with eigenvalues of $+1$ and $-1$,
respectively. If these labels had been adjacent in the
cycle of possible eigenvalues ($1$, $i$, $-1$, $-i$, ...),
a Chern transfer $\Delta C=\pm1$ would have been expected; but
because they are not, we get $\Delta C=\pm2$.~\cite{Fang:2012}

%---------------
% singular point ---> small band gap
%----------------
Finally, we note that since the minimum avoided crossings are
not in general located on the high-symmetry X-M or $\Gamma$-M
$k$-path when octahedral rotations are present, the actual band gap is 
smaller than the one obtained from a band structure plotted along
the $\Gamma$-X-M-$\Gamma$ high-symmetry lines.
For example, a direct inspection of Fig.~\ref{fig:modelb}(a)
suggests a gap of $30$\,meV, compared to the true value of $25$\,meV
obtained from a more careful scan over the full 2D BZ.

%%%%%%%%%%%%%%%%%%%%%%%%%%%%%%%%%%%%%%%%%%%%%%%%%%
\section{first-principles calculations}
%%%%%%%%%%%%%%%%%%%%%%%%%%%%%%%%%%%%%%%%%%%%%%%%%%

%------------------------------
%Fig3: 1st principles bands
%------------------------------
\begin{figure*}
\includegraphics[width=18.0cm]{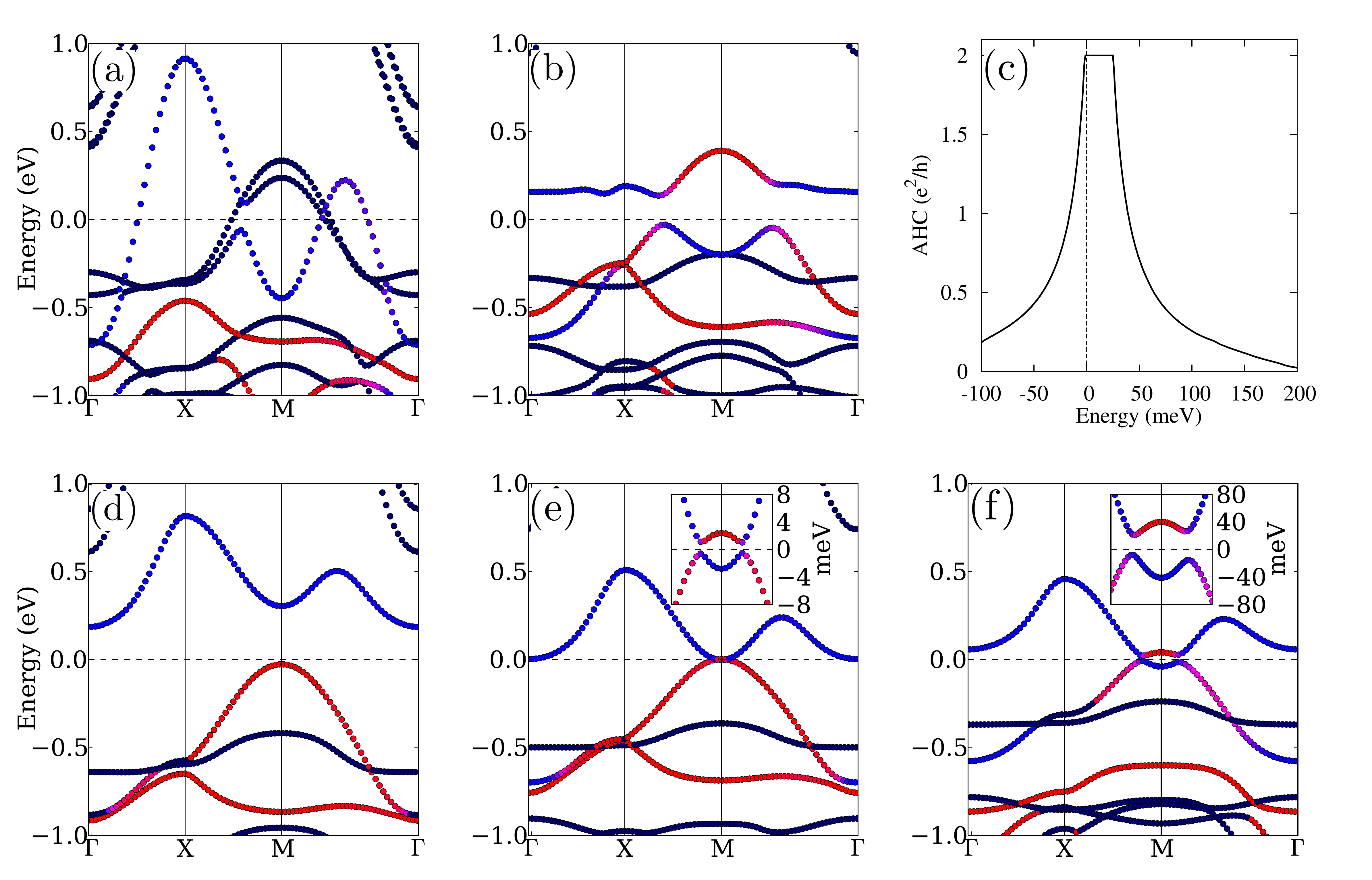}
\caption{(Color online)
Electronic structures of the LMIO/LAO superlattices.
Red (blue) color-coding highlights the character of  the $d_{z^2}$ ($d_{x^2\!-\!y^2}$)
orbitals of the Mn atoms, and black indicates the Ir-$5d$ states.
All calculations are done with $U\!=\!5.0$\,eV and $J\!=\!1.0$\,eV on Mn
unless stated otherwise.
(a) Hypothetical structure without octahedral rotations (see main text for details).
(b) Hypothetical structure with $15^\circ$ rotations about the $z$-axis
in the LMIO layers only.
(c) Calculated anomalous Hall conductivity for case (b).
(d) Relaxed structure at zero epitaxial strain.
(e) Relaxed structures at $2\%$ tensile epitaxial strain.
(f) Same as (e) but with $U$ on Mn sites increased to $7.0$\,eV. 
Insets in (e-f) zoom in on the region around the M point.
Dashed lines denote the Fermi energies.
}\label{fig:bands}
\end{figure*}

In this section we demonstrate how the tight-binding model
discussed above can be realized in more realistic systems.
The spin-polarized half-filled $e_g$ states could be realized by
a $d^4$ or $d^9$ configuration, while non-spin-polarized sublattice
could be populated by $d^0$,
$d^6$, or $d^{10}$ ions.
In this work we considered a specific
system consisting of a monolayer of LMIO sandwiched into an LAO
environment, as shown in Fig.~\ref{fig:model}(a), although the
realization of the
tight-binding model is not limited to this specific system.
We have chosen LAO as the host environment because it has a large bulk gap
of $5.6$\,eV, so that the states around the Fermi energy ($E_F$) will be dominated by
the orbitals in the LMIO layers.

%---------------
% numerics
%----------------
Our first-principles calculations are done using the projector
augmented wave method as implemented in VASP.~\cite{vasp}
The exchange-correlation potentials are approximated using the
Perdew-Burke-Ernzerhof functionals.~\cite{Perdew:1996}
For all the structures considered, the in-plane lattice constants
are fixed at $3.789$\,\AA, the cubic lattice constant of bulk LAO.
For the self-consistent total-energy calculations,
the plane-wave energy cutoff is taken to be $500$\,eV.
All our calculations are carried out using the
$\sqrt{2}\!\times\!\sqrt{2}\!\times\!3$ supercell shown
in Fig.~\ref{fig:model}(a), which also accommodates octahedral
rotations about the $z$ axis, and a k-point set corresponding to
an $8\!\times\!8\!\times\!4$ mesh in the full BZ
is used.

To treat Coulomb interactions for open shells,
we applied the GGA+U
method\cite{Lichtenstein:1995} with double-counting considered
in the fully localized limit.
Since the $t_{2g}$ shells of Ir
are almost fully occupied, the GGA+U corrections are only applied to
Mn sites.  Initially our calculations are all carried out with
$U\!=\!5.0$\,eV and
$J\!=\!1.0$\,eV,~\cite{Pavarini:2010} corresponding to
commonly accepted values for Mn$^{3+}$.  Later, we study the
effect of varying the $U$ value on the Mn sites, as
discussed below.
In all our calculations, we assume that the magnetic
moments of Mn are ferromagnetically coupled.
To shift the $4f$ states of La away from $E_F$, we impose
$U_{4f}\!=\!11$\,eV and $J_{4f}\!=\!0.68$\,eV as used for other
calculations on heterostructures.~\cite{Okamoto:2006}
The AHC is obtained by Wannier interpolation using an
effective Hamiltonian constructed in a basis of $128$ maximally localized
Wannier functions~\cite{wannier} corresponding to all Mn-$3d$, Ir-$5d$,
and O-$2p$ orbitals in the supercell.

%---------------
% hypothetical structure without rotation
%----------------
Consider first a hypothetical structure without octahedral rotations,
specifically one in which
the in-plane Mn-O and Ir-O distances are set
to be equal, and the out-of-plane Mn-O distance is set
to be $2.0$\,\AA. 
The first-principles band structure is shown in Fig.~\ref{fig:bands}(a).
Due to the strong atomic SOC of Ir atoms, their $t_{2g}$ bands are separated into a group
of four lower-lying $J\!=\!3/2$ bands and two higher $J\!=\!1/2$
bands.~\cite{Kim:2008}
The bands around $E_F$ are mostly a mixture of $e_g$ bands
from Mn and the $J\!=\!1/2$ bands from Ir.
The $e_g$ states of Mn are
half-filled, leading to an atomic magnetic moment of about
$4\,\mu_\mathrm{B}$.
The hybridization of the Ir $t_{2g}$ states with the Mn $e_g$
states induces small (about $0.05\,\mu_\mathrm{B}$) magnetic moments on
the Ir sites.

%---------------
% hypothetical structure with 15 degree rotation
%----------------
Introducing octahedral rotations in the LMIO layers leads to significant
changes of the band structure by inducing additional hybridizations. 
Fig.~\ref{fig:bands}(b) shows the band
structure with a staggered rotation of the MnO$_6$ and
IrO$_6$ octahedra of $15^\circ$ about the $z$ axis, while all the other
degrees of freedom remain fixed.  Now only two bands, mainly of
Mn $d_{x^2-y^2}$ and $d_{z^2}$ character, are left around $E_F$,
although these orbitals hybridize strongly with the Ir $5d$ orbitals.
Recalling the arguments given above in connection with our
tight-binding model, such hybridization is crucial for inducing the
effective SOC $\lambda^{(2)}$ in Eq.~(\ref{eq:model}),
which in turn helps to give a nontrivial gap.  We confirm that the gap is
indeed nontrivial, with a quantized AHC of $2$\,$e^2/h$,
by direct calculation as shown in Fig.~\ref{fig:bands}(c).
The topological band inversion is also evident in the band
structure of Fig.~\ref{fig:bands}(b), where the band characters have
clearly exchanged between the conduction and
valence bands around the M point.
We note that the size of the gap is about $26$\,meV as measured by the
width of the quantized AHC plateau, which
is smaller than the gap obtained from inspection along the
high-symmetry $k$-path in Fig.~\ref{fig:bands}(b); this is
again due to the fact that the avoided-crossing points are not located on
the high-symmetry lines (cf.~Fig.~\ref{fig:modelb}(d)).

%---------------
% structural variations upon relaxation
%----------------
To be more realistic, we relaxed the structures by allowing the
out-of-plane lattice constant and internal coordinates to vary,
but keeping the in-plane lattice constants fixed at those of LAO.  
We find the relaxed octahedral rotation angle in the LMIO layers
to be $15.6^\circ$, and the relaxed out-of-plane Mn-O distance
is about $2.02$\,\AA.  By these measures,
the ideal structure discussed above is quite reasonable. 
However, the most drastic change occurs locally in the MnO$_6$ 
octahedra, where the local $c/a$ ratio (i.e., the ratio of apical
to in-plane Mn-O bond lengths) increases to $1.06$, from $1.02$ 
in the ideal structure.  
This change results from a contraction of the in-plane Mn-O distances.

Fig.~\ref{fig:bands}(d) shows the band structure for the fully relaxed
structure. Evidently the $d_{x^2-y^2}$ bands are shifted to higher energies
due to the variations of the on-site energies of the $d_{x^2-y^2}$ 
and $d_{z^2}$ orbitals caused by the local distortions of the MnO$_6$
octahedra.
The resulting $d_{x^2-y^2}$ and $d_{z^2}$ bands no longer overlap
anywhere in the BZ, and as a result the gap at $E_F$ is topologically
trivial, as verified by our calculations of the AHC (not shown).
Another consequence of the structural relaxations is that the local
conduction-band minimum at $\Gamma$ has shifted downward and now
falls about $120$\,meV below the conduction-band minimum at M.
This is caused by
a change in the sign of the hopping parameter between $d_{x^2-y^2}$
orbitals located on NN sites,~{\it i.e.},~$t_{1a}$ in Eq.~(\ref{eq:model}).
Thus, even if some means could be found to restore the band inversion
at M, this reversal in the energy ordering of the conduction-band
minima could prevent the maintenance of a global gap, forcing the
system metallic.

%---------
% tuning by strain
%---------

To overcome these negative effects of the structural relaxations,
which disfavor the topological
phase, tensile epitaxial strain can be applied to increase the in-plane
lattice constants and decrease the out-of-plane one, thus reducing
the local octahedral distortions.
Fig.~\ref{fig:bands}(e) shows the band structure with a $2\%$
tensile epitaxial strain applied to the LAO substrate.
In this case the $d_{x^2-y^2}$ bands are shifted downward in
energy relative to the $d_{z^2}$ bands, once again overlapping with
them.  The gap opened around M shows a typical
anticrossing behavior, as emphasized in the inset of
Fig.~\ref{fig:bands}(e), and our calculation of the AHC confirms that it
is topologically nontrivial.
However, the magnitude of the gap is quite small, only about 1\,meV.
This is a consequence of the fact that the off-diagonal terms in the 
Hamiltonian of Eq.~(\ref{eq:model}) vanish as one approaches the
M point because of the form of $g_1(\mathbf{k})$ and $g_2(\mathbf{k})$.
To our satisfaction, we observe that the $d_{x^2-y^2}$ conduction-band
minimum at $\Gamma$ remains above that at the M point, if only barely
(by $\sim\!5$\,meV), so that the gap around M is a true global gap.
We conclude that a tensile strain of at least $2\%$ is needed to
obtain the QAH state, and speculate that out-of-plane uniaxial 
pressure could help further.

%---------------
% effect of e-e correlations
%----------------
Interestingly, increasing the strength of the local Hubbard $U$ on
the Mn sites also tends to stabilize the topological phase.
Fig.~\ref{fig:bands}(f) shows the resulting band structure obtained by 
increasing the Hubbard
parameter to $U\!=\!7$\,eV on the Mn sites, with the conditions otherwise
the same as in Fig.~\ref{fig:bands}(e) (\textit{i.e.,} relaxed with
$2\%$ tensile epitaxial strain).
Larger $U$ not only shifts the conduction-band minimum at $\Gamma$
upwards away from $E_F$, but also enhances the magnitude
of the nontrivial topological gap opened around the M point.
The magnitude of the global band gap is calculated to be about $4$\,meV.
This is much smaller than the
band gap derived from states along the high-symmetry $k$-path,
which is about $25$\,meV, again because
the avoided crossings are not located on the high-symmetry lines as
explained in Sec.~II.
Moreover, significant changes occur in the hybridizations between the 
valence states, caused by the enhanced local atomic $U$ values on 
the Mn sites. 
For instance, the $d_{z^2}$ bands are shifted to lower energies 
(Fig.~\ref{fig:bands}(e) versus Fig.~\ref{fig:bands}(f)),
and the first valence band below $E_F$ acquires more $d_{x^2-y^2}$ character
with increasing $U$ because of more significant hybridization with the $d$ states of Ir atoms.

%%%%%%%%%%%%%%%%%%%%%%%%%%%%%%%%%%%%%%%%%%%%%%%%%%
\section{Conclusions}
%%%%%%%%%%%%%%%%%%%%%%%%%%%%%%%%%%%%%%%%%%%%%%%%%%

In conclusion, we have demonstrated the possibility of achieving
a quantum anomalous Hall phase on a square lattice via
an appropriate pattern of 
intersite spin-orbit couplings between $d$ orbitals, which
can be realized in double-perovskite monolayers.
We have shown that for a half-filled manifold of $e_g$ orbitals,
an effective SOC can be induced by hybridizing
with other $d$ orbitals located on the neighboring sites,
even though no direct on-site SOC is present.
We have found, in particular, that octahedral rotations can
induce an effective SOC between $e_g$ orbitals
located on NNN sites.  We have demonstrated 
that a simple tight-binding Hamiltonian
encoding the most important features of the interactions gives
rise quite generically to a quantum anomalous Hall phase.
Then, using first-principles calculations, we have also shown 
that such a model can be realized in La$_2$MnIrO$_6$ monolayers.
The gap can be as large as $26$\,meV in the ideal case.
However, there are several open issues that need further
investigation for this system.
First, we have assumed 
ferromagnetic order, even though there is some tendency of the 
magnetic moments of the Mn ions to be coupled 
antiferromagnetically. This problem may be
remedied by choosing a substrate with a magnetic order 
that can enforce the desired ferromagnetic state.
Second, there is the issue of the assumed checkerboard compositional
order. Even though both Mn and Ir are $3+$ ions, which by itself
would give no strong tendency toward ordering within
the La$_2$MnIrO$_6$ monolayer, we argue that such a tendency may
come instead from the substantial difference in the ionic radii.
Third, it would be useful to understand the role of correlations 
in more detail.
We have found that enhancing the local Hubbard $U$ favors
larger band gaps while maintaining the topological non-triviality,
but the interplay of electronic correlations and 
SOC on the square lattices deserves further investigation from
beyond-DFT methods.

Finally, as discussed above,
the global gap may not remain open after the contraction of 
the in-plane Mn-O distances with relaxation.
We have shown that this problem may be overcome by engineering structures
utilizing tensile epitaxial strain of about $2\%$, making use of 
the sensitivity of the relevant $d_{x^2-y^2}$ and $d_{z^2}$ bands
to local distortions. Further stabilization of 
the QAH phase might be achieved 
by  varying the choice of the inert perovskite surrounding material,
by applying vertical uniaxial pressure in addition to the tensile 
epitaxial strain,
or by chemical doping within the double-perovskite layer or in
the surrounding material.
Other combinations of transition-metal ions, in which one has
a half-filled $e_g$ shell, should also be explored.
Lastly, we suspect that the idea of intersite SOC on the square
lattices should be applicable to ions with partially filled $t_{2g}$
shells as well.
These interesting questions are left for future investigations.

\acknowledgments
This work was supported by NSF Grant DMREF-12-33349.

\end{document}